# Estimating See and Be Seen Performance with an Airborne Visual Acquisition Model

Ngaire Underhill, Evan Maki, Bilal Gill, and Andrew Weinert

**Abstract** — Separation provision and collision avoidance to avoid other air traffic are fundamental components of the layered conflict management system to ensure safe and efficient operations. Pilots have visual-based separation responsibilities to see and be seen to maintain separation between aircraft. To safely integrate into the airspace, drones should be required to have a minimum level of performance based on the safety achieved as baselined by crewed aircraft seen and be seen interactions. Drone interactions with crewed aircraft should not be more hazardous than interactions between traditional aviation aircraft. Accordingly, there is need for a methodology to design and evaluate detect and avoid systems, to be equipped by drones to mitigate the risk of a midair collision, where the methodology explicitly addresses, both semantically and mathematically, the appropriate operating rules associated with see and be seen. In response, we simulated how onboard pilots safely operate through see and be seen interactions using an updated visual acquisition model that was originally developed by J.W. Andrews decades ago. Monte Carlo simulations were representative two aircraft flying under visual flight rules and results were analyzed with respect to drone detect and avoid performance standards.

## I. Introduction

THE National Airspace System (NAS) is a complex and evolving system that enables safe and efficient aviation. Drones are leveraging the NAS in different ways than existing traditional aviation. To enable growth in the industry and new economic opportunities, drones must integrate into the NAS without degrading overall safety or efficiency. To help achieve this, standards developing organizations have established and are updating Detect and Avoid (DAA) performance standards.

DAA systems are designed as an electronic means of compliance to the primarily visual-based separation responsibilities of a pilot and to comply with applicable operating rules of Title 14 of the Code of Federal Regulations (14 CFR). Some of these rules include 14 CFR §91.3, .111, .113(b), .115, .123, .181(b) and Part 107.37, which prescribe that aircraft must not operate carelessly or recklessly; not operate so close to another aircraft so as to create a Mid Air Collision (MAC) hazard; see and avoid other aircraft; and give way to other aircraft that have the right of way. Just like piloted aircraft, a drone yielding to other aircraft with the right of way should result in aircraft having sufficient separation between them as to not interfere with the right of way of the other aircraft. Aircraft that have the right-of-way should not need to maneuver with respect to another aircraft in order to maintain safe flight. If avoidance maneuvers are needed to mitigate an imminent collision hazard, then that is an indication that right-of-way was violated and the aircraft without the right-of-way did not adequately yield to be sufficiently well clear.

This work was supported by Sabrina Saunders-Hodge, Adam Hendrickson, and John Miller from the Federal Aviation Administration under Air Force Contract No. FA8702-15-D-0001". DISTRIBUTION STATEMENT A. Approved for public release. Distribution is unlimited. This material is based upon work supported by the Federal Aviation Administration under Air Force Contract No. FA8702-15-D-0001. Any opinions, findings, conclusions or recommendations expressed in this material are those of the author(s) and do not necessarily reflect the views of the Federal Aviation Administration. This document is derived from work done for the FAA (and possibly others), it is not the direct product of work done for the FAA. The information provided herein may include content supplied by third parties. Although the data and information contained herein has been produced or processed from sources believed to be reliable, the Federal Aviation Administration makes no warranty, expressed or implied, regarding the accuracy, adequacy, completeness, legality, reliability, or usefulness of any information, conclusions or recommendations provided herein. Distribution of the information contained herein does not constitute an endorsement or warranty of the data or information provided herein by the Federal Aviation Administration or the U.S. Department of Transportation. Neither the Federal Aviation Administration nor the U.S. Department of Transportation shall be held liable for any improper or incorrect use of the information contained herein and assumes no responsibility for anyone's use of the information. The Federal Aviation Administration and U.S. Department of Transportation shall not be liable for any claim for any loss, harm, or other damages arising from access to or use of data information, including without limitation any direct, indirect, incidental, exemplary, special or consequential damages, even if advised of the possibility of such damages. The Federal Aviation Administration shall not be liable for any decision made or action taken, in reliance on the information contained herein.

N. Underhill is with the MIT Lincoln Laboratory, Lexington, MA 02421 USA (e-mail: ngaire.underhill@ll.mit.edu).

E. Maki is with the MIT Lincoln Laboratory, Lexington, MA 02421 USA (e-mail: evan.maki@ll.mit.edu).

B. Gill, was with MIT Lincoln Laboratory, Lexington, MA 02421 USA. He is now with PickNik Robotics, Boulder, CO 80301.

A. Weinert is with the MIT Lincoln Laboratory, Lexington, MA 02421 USA (e-mail: andrew.weinert@ll.mit.edu).

According to the International Civil Aviation Organization (ICAO), safe operations are supported through conflict management consisting of three layers [1]: strategic conflict management through airspace organization and management, demand and capacity balancing, and traffic synchronization; separation provision; and collision avoidance. Separation provision is the tactical process of keeping aircraft away from hazards by at least the appropriate separation minima; while collision avoidance is the process of avoiding immediate proximity hazards. Thus, to satisfy the FAA defined separation responsibilities, a DAA system, using the ICAO conflict management concepts, must have separation provision and collision avoidance capabilities. The responsibility to maintain separation is always present regardless of air traffic density, estimated airspace collision rate, operating regulations, or minimum safe altitudes.

*A. Motivation and Scope*

To develop performance-based standards that satisfy applicable operating rules, there is need for a methodology to design and evaluate DAA systems, where the methodology explicitly addresses, both semantically and mathematically, the appropriate operating rules. Accordingly, the methodology should not rely solely on estimated encounter rates, since neither 14 CFR §91.113 nor §107.37 depend on them. Rather, there is a need to demonstrate how DAA requirements are a safe alternate means of compliance to the rules.

An adequate DAA system will not only satisfy 14 CFR 91.113, but also satisfy the underlying see-and-be-seen safety concept that it is built upon in order to maintain safe interactions. Hence, an adequate DAA system will be one where the total interaction risk between aircraft is assessed (e.g. see-and-be-seen interaction) rather than only a direct comparison of DAA to see-and-avoid performance. This will ensure that DAA interactions are compatible and as safe as existing aircraft-aircraft interactions.

Our scope was informed by the DAA performance standards for smaller drones: ASTM F3442/F3442M-20 [2] and the RTCA DO-396 minimum operating performance standard (MOPS) for ACAS sXu [3]. The scope was also informed by the needs and synergic research of various FAA branches We only considered scenarios where air traffic control (ATC) is not providing separation services. We did not consider strategic conflict management nor the terminal airport environment. Out of scope were also performance requirements not explicitly related to DAA, such as system robustness, availability, maintenance, noise, assurance, or reliability.

*B. Objectives and Contributions*

Our primary objective was to simulate how onboard pilots safely operate through "see and be seen" interactions with other aircraft. Meeting this objective supports a larger effort to propose and use a methodology to derive DAA performance requirements for low altitude drone operations subject to FAA regulations. The methodology was based on concepts established by the U.S. Army in support of their Ground Based Sense and Avoid (GBSAA) system that has been operating in the NAS since 2016 [4], [5]. Based on a quantitative estimation of these "see and be seen" interactions, we seek to derive safety performance requirements by estimating the probability of a MAC given an encounter between aircraft, based on aircraft size which affects both visual acquisition and the likelihood of a MAC.

To model the "see and be seen" behavior of conventional crewed aircraft, the primary contribution was an updated model of visual acquisition originally developed by J.W. Andrews [6]–[9]. This model was integrated into a Monte Carlo simulation and with an empirical, rule-based stochastic pilot response model [10]. Using this simulation, the performance by onboard pilots to minimize the risk of near midair collisions (NMACs) was estimated for different scenarios. We then discuss if the DAA NMAC safety performance targets defined in the ASTM F3442/F3442M-20 standard would enable an equivalent or better level of safe aircraft-aircraft interaction as simulated "see and be seen" scenarios. Additional contributions are the release, under a permissive open-source license, of the updated visual acquisition model, end-to-end simulation, and dataset of simulated encounters. All discussed simulations and results should be repeatable by third parties.

II. FOUNDATIONAL MODEL OF VISUAL ACQUISITION

In the 1980-1990s, J.W. Andrews at MIT Lincoln Laboratory (MIT LL) led development of a mathematical model of air-to-air visual acquisition of an aircraft by a human pilot in an aircraft cabin. The model was notably used to analyze the 1986 Cerritos midair collision between a large fixed-wing multi-engine and a small fixed-wing single engine aircraft [6]. It models acquisition, under daylight conditions, as a nonhomogeneous Poisson process. It assesses probability, not spatial estimation, of visual acquisition: the position uncertainty of the surveilled aircraft is not modeled.

The model uses Koschmieder's Law to model how the contrast of the target is reduced when the atmosphere is not clear. This relationship is based on studies that determined that the probability of sighting an aircraft is related to the product of the visual angle subtended by the target area and the target's contrast against its background. In this equation, the opportunity for visual acquisition is the sum of the opportunities given a search effectiveness β, target area A, atmospheric visibility R, and range between aircraft r. Equation 1 defines this visual acquisition rate, λ, as

$$\lambda = \beta \frac{A}{r^2} exp\left(\frac{-2.996r}{R}\right) \qquad (1)$$

Of these variables, R defines the atmospheric environment as the prevailing visual range along the line of sight to the target aircraft when aircraft is clear of clouds. The target area is the size of the aircraft as viewed from the cabin; it is dependent on the relative geometry and the two-dimensional projection of the aircraft. As aircraft fly closer, the target area increases.

Search effectiveness β is a curve-fitted parameter, derived from flight test data, that accounts for all the human performance effects when visually acquiring aircraft. It is modeled as the rate of visual acquisition per solid angle of target size per second of search. It accounts for the inherent contrast of the aircraft in the sky but is independent of prevailing atmospheric visibility. While search effectiveness β is often described as a constant, the implementation by J.W. Andrews denoted an effective β that can transition between different values. This transition can be due to transitioning in and out of a pilot's field of view (FOV) or due to an alert to the pilot about nearby traffic. This implementation also assumed the visual acuity (resolution limit) of the human eye is one arc minute. Other factors influence search effectiveness, including: phase of flight, intruder location, relative intruder movement, pilot workload, and pilot prioritization of situational awareness.

The derived search effectiveness β value is wholly dependent upon flight tests; with J.W. Andrews emphasizing that "The model has a significant limitation: it can be applied only where the pilot performance level can be assumed to approximate that for which flight test data are available [6]." The flight test conditions described in [8], resulted in the original implementation [7] that assumed the exclusion of unusual or non-standard visual conditions and assumed a non-accelerated collision course. Because of the target area projection, the target aircraft have only small bank and pitch angles. It also was not usable when a target remained at the visual resolution threshold for long periods of time, could not be used to resolve field of view obstructions, and could not be used to produce conclusions based on the rapid acquisition probabilities immediately prior to a MAC.

Clothier et al. [11] concluded that based on the assumptions and flight test conditions in [6], [8], the model is likely to be overly optimistic of human detection performance; but still leveraged the model to propose performance requirements as it has been broadly considered state of the art. Santel [12] also noted the model's dependence on flight test results and assessed if the model parameters derived from the J.W. Andrews directed flight tests were applicable to different encounters. Santel makes a similar conclusion as Clotheier et al, that "investigators came to suspect that the visual attention demonstrated by the accident flight crews may have been notably lower than the effectiveness demonstrated by Andrews' flight test participants." Additionally, Carreño et al. used the model to design and evaluate a prototype assistive technology to enhance pilot's ability to see and avoid nearby traffic [13]; Woo et al. expanded the model for visual acquisition of smaller drones[14]; and James et al. used it as a reference to evaluate a deep convolutional neural network to detect aircraft using machine vision[15].

### III. ADAPTION FOR RESPONSIVE SIMULATION

The reference algorithm and implementation for the visual acquisition model are defined in [7] and provided a post-hoc calculation of the probability of visual acquisition of an aircraft during an encounter. The target aircraft had to have a consistent bearing, closure rate, and visual area; both aircraft were therefore restricted to flying straight, consistent trajectories: changes in heading (turns), altitude (climbs/descents), and speed (acceleration/deceleration) were not allowed. These original constraints significantly limited aspects desired for rigorous and diverse scenario evaluation. In response, we implemented the model in Simulink for the DAA Evaluation of Guidance, Alerting, and Surveillance (DEGAS) framework [16] and also modified the integral calculation, which evaluated the whole trajectory, to calculate incremental contributions to the visual acquisition probability. This enabled evaluation at each step through the encounter as it progressed to determine if visual acquisition would occur; rather than evaluate post-hoc if the target aircraft was visually acquired at any timestep.

Using DEGAS, an aircraft's state can vary each timestep and allow for changes to speed, altitude, and heading. Unlike like [7], aircraft can be simulated to literally "see and avoid" each other and react based on the cumulative probability of visual acquisition and aircraft states at each timestep. The DEGAS implementation of the model extends that probability in greater detail since an encounter that is run a second time could still be visually acquired (result in the same outcome for the original implementation) but under this new implementation the time of acquisition can also vary. DEGAS also enables simulating encounters repeatable with different surveillance models instead of visual acquisition, which enables the "see and be seen" results be a baseline to be compared against. For an encounter between two aircraft, either one or both aircraft could react and deviate from their initially planned trajectories to increase separation.

Furthermore, the original [7] implementation evaluated the full encounter at once and incorporated alerted search if a DAA system was alerting/providing information about the intruder. In the DEGAS implementation, the alerted search functionality is present but turned off in the current implementation to solely enable visual acquisition based "see and avoid" results. Although visual acquisition is still applicable with alerted search and a potential subject of future work. Target acquisition is also different: [7] calculated whether visual acquisition occurred during an encounter or not, however, the DEGAS implementation allows periods before visual acquisition, periods where the target has been visually acquired, and periods when the target has left the FOV and therefore visual acquisition (and therefore intruder position/movement) is no longer known as it was in [7]. Related metrics calculated in DEGAS include: what time was the intruder visually acquired, what were the relative aircraft geometries when acquisition occurred, and how long had the intruder been within visual acquisition range before acquisition.

To model "see and be seen," where each aircraft can avoid the other, the DEGAS implementation was designed to an operational concept of if an aircraft has been seen, the pilot will continue to track the aircraft until it is no longer in their FOV. The response maneuvers consisted of an empirical, rule-based stochastic pilot response model [10] that reacted to the Detect and AvoID Alerting Logic for Unmanned System (DAIDALUS) [17], a reference implementation of the alerting and guidance functional requirements described in Appendix G of the MOPS Phase I for UAS developed by RTCA SC-228. We do not assume DAIDALUS, as a computer algorithm, is representative of decision making by human pilots. Additionally, DAIDALUS is provided perfect state and velocity information, another overly optimistic expectation on human performance where range and speed of other aircraft are visually inferred rather than exactly known. We assume that results using this simulation are overly optimistic of human detection performance and pilot avoidance decisions, aligning with assumptions established in the literature. DAIDALUS was selected because it was easily and freely available and no appropriate model of pilot tactical decision making was available. This was acceptable because of our objective to validate and derive safety performance targets for new entrants. By overestimating human performance, any conclusions would be risk adverse and not lead to DAA performance requirements that are less safe than current interactions between aircraft.

IV. MODEL IMPROVEMENTS AND NEW FUNCTIONALITY

In addition to adapting the visual acquisition model for responsive simulation, we enhanced the Simulink model to enable new scenarios and to apply realistic constraints that were previously not included. We summarize these enhancements here; please refer to the released software for more details.

These updates enabled the model to better account for maneuvering targets with changes to heading, altitude, and speed. The model additionally can handle all bank and pitch angles accurately. However, many limitations still remain, including not accounting for unusual visual conditions. Additionally, encounters where the target is at the visual limit for long periods of time are not accurately representative and conclusions cannot be made based on visual acquisitions made immediately before collision.

### A. Aircraft Target Area Measurement Improvements
*1) Expansion of the target aircraft dataset*

The original implementation of the model utilized seven aircraft shapes. This covered a small percentage of aircraft types, sizes, and shapes. We increased the number of aircraft shapes in the model to better represent aircraft diversity and enable specific aircraft frame testing during simulation. We utilized a database of wireframe aircraft models, expanding the availability diversity and accuracy by enabling direct selection of specific target aircraft.

*2) Utilizing Wireframe Models for Aircraft Area*

177 wireframe models were used to calculate the visible surface area of the aircraft from different angles. [7] projected eight aircraft silhouettes onto a two-dimensional surface where the individual pitch/lateral and roll/longitudinal axis projections were fractionally combined to produce a rough estimate of visible projected area. Projecting the wireframe models from a variety of angles resulted in more accurate estimations of the aircraft surface area; particularly for non-small pitch and bank angles – removing another limitation of [7]. The use of wireframe models was motivated by [18] and [19]; and the calculated projected areas are available under a permissive open source license [20]. Table 1 are example projected areas for a Cessna 172 where azimuth = 90 degrees and elevation = 0 degrees are the side of the fuselage. Figure 1 illustrates the projection of a 90 degrees azimuth and 0 degrees elevation orientation.

TABLE I
EXAMPLE CESSNA 172 PROJECTED AREAS

| Azimuth (deg) | Elevation (deg) | Area (ft$^2$) |
|---|---|---|
| 0 | 0 | 110 |
| 0 | -90 | 430 |
| 0 | 90 | 430 |
| 0 | -75 | 407 |
| 0 | 75 | 424 |
| 0 | -60 | 363 |
| 0 | 60 | 388 |
| 60 | 0 | 143 |
| 75 | 0 | 145 |
| 90 | 0 | 118 |

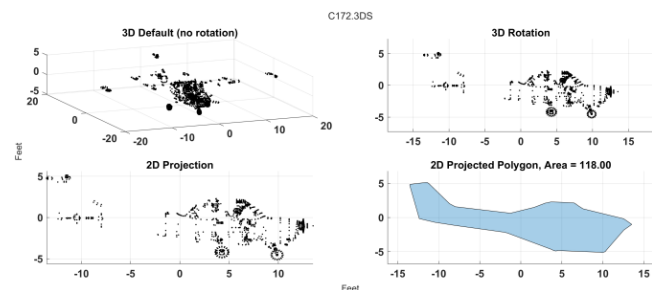

Figure 1. Cessna 172 3D to 2D projection for a 90 degrees azimuth and 0 degrees elevation.

### B. Field of View

In [7], only left and right FOV limits were implemented. While FOV is aircraft specific, we implemented, in DEGAS, limitations on up, down, left, and right angular view based on the FAA published AC 25.773-1 - pilot compartment view design considerations. Implementing FOV introduces new conditions where one aircraft could never acquire the other as well as the potential to lose visual acquisition since an aircraft could fly out of FOV as well. The DEGAS implementation enabled a loss or prevention of visual acquisition, based on FOV limitations. Table 2 summarizes the FOV values

TABLE 2
FIELD OF VIEW VALUES

| FOV | DEGAS [20] | ATC-151 [7] |
|---|---|---|
| Up | 15 degrees | 90 degrees (implicit) |
| Down | 17 degrees | 90 degrees (implicit) |

| | | | |
|---|---|---|---|
| **Left** | 120 degrees | 120 (explicit) | degrees |
| **Right** | 80 degrees | 90 (explicit) | degrees |

### C. Direction of View (DOV)

Another realistic aspect of visual acquisition is the pilot scan technique. [7] did not specifically represent how a pilot would scan the sky to look for intruder aircraft. Direction of view (DOV) models the scan technique used by pilots where their attention and ability to visually acquire an intruder aircraft would rotate between different sections of sky. DOV in DEGAS can be configured to prioritize different parts of the FOV and prioritize dwelling a specific direction for increased attention

### V. RESULTS

Using DEGAS, we simulated pairwise encounters of two aircraft flying VFR in "see and be seen" interactions; both aircraft could maneuver to increase separation. We assessed performance using a standard safety performance metric widely used in collision avoidance standards and assessed how performance is dependent on different model parameters.

### A. Pairwise Encounters

Two encounter sets were created using the encounter sampling process described in [21] and statistical models of aircraft behavior described in [22] and sourced from [v1.3]. One encounter set was created to represent fixed wing intruder aircraft encounters, and the other was created for rotary wing aircraft encounters. Encounters had a 220 seconds duration, with a time of closest approach (CPA) at 180 seconds. Initial lateral separation was at least 3.5 nautical miles and there was no restriction on initial vertical separation. The unmitigated horizontal miss distance (HMD) and vertical miss distance (VMD) distributions were importance sampled to better assess rare scenarios of a loss of separation between aircraft. The edges of the proposed HMD distribution [-2000, -500, 500, 2000] feet and the proposed VMD edges were [-450, -100, 100, 450] feet. Aircraft could have any altitude of [200, 4000] feet above ground level and speeds of [60, 250] knots.

### B. Model Parameters

We simulated the encounters with different configurations of the visual acquisition model. We prioritized enhancing and testing the surveillance model and iterating over its surveillance parameters to validate it against the original model and explore the performance for a variety of different scenario conditions. Foremost, the set of atmospheric range was [2, 3, 4, 5] nautical miles, and the set of search effectiveness was [4250, 8500, 12500, 17000]. Note 17000 was the search effectiveness of unalerted search derived by and solely used by J.W. Andrews [8]; while this analysis assessed the performance sensitivity to different search effectiveness values. Both aircraft had the same search effectiveness. Maneuvers commenced 10 seconds after visual acquisition to account for total human response latency and aircraft response lag delay; this delay aligns with concepts described in AC 90-48D [23] informed the two DOV configurations. These configurations were weighted dwell time, defined in [23], and uniformly weighted dwell times across DOV partitions. Weights in Table 2 are normalized when simulated. FOV and DOV are not symmetric because pilot seats are not centered on an aircraft and the aircraft is piloted from the left seat.

TABLE 3
DOV WEIGHTED DWELL TIMES INFORMED BY [23].

| Partition | DOV Partition | Weight |
|---|---|---|
| Left | -120 to -60 | 3.5 |
| Left Center | -60 to 0 | 5.75 |
| Right Center | 0 to 60 | 5.75 |
| Right | 60 to 90 | 3.5 |

### C. Risk Ratio Performance Metric

Risk ratio was the standard performance metric used to determine the effectiveness of pairwise visual acquisition and avoidance in lowering the probability of an NMAC.

$$Risk\ Ratio = \frac{P(NMAC\ |Visual\ Aquisition\ \&\ Maneuver)}{P(NMAC\ |\ nominal\ encounters)}$$

The denominator is calculated by simulating nominal encounters without maneuvers in response to visual acquisition and calculating the weighted number of NMACs in the encounter set. The numerator is the weighted quantity of NMACS when simulating the pairwise encounters using the visual acquisition model and aircraft response (from DAIDALUS and the pilot model) with the objective of increasing separation between aircraft. Weighted values are used because of the importance sampling. We calculated unresolved and induced risk ratios by separating the mitigated NMACs into two categories. Unresolved risk ratio is calculated using NMACs that are also present in the nominal scenario (denominator) and where "see and be seen" did not provide sufficient mitigation. Induced risk ratio is calculated using NMACs that are not present in the denominator – these are NMACs that are created by the addition of the mitigations. Some induced NMACs are expected due to the dynamic encounter situation, but excessive amounts should prompt concern and could indicate the logic is creating a hazard under specific conditions.

### D. Results and Visualization

Simulation results are visualized as bar graphs organized into unresolved, induced, and the summation, total risk ratio, for a variety of parameter and encounter combinations. Figure 2 visualizes encounters between fixed-wing single engine aircraft with an unweighted (uniform) dwell times and Figure 3 is the same fixed-wing single engine aircraft with the weighted DOV distribution defined in Table 2. Figures 4 and 5 are the analogous results for encounters between rotorcraft.

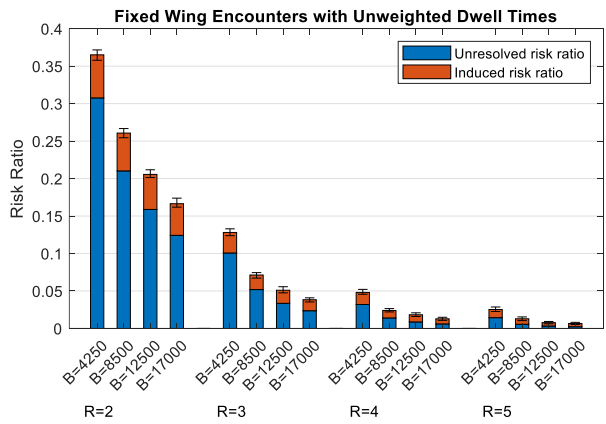

Figure 2. Fixed Wing Encounter Set with Unweighted Dwell Times

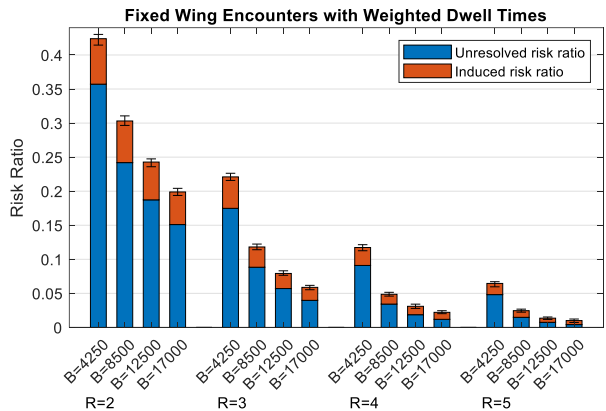

Figure 3. Fixed Wing Encounter Set with Weighed Dwell Times

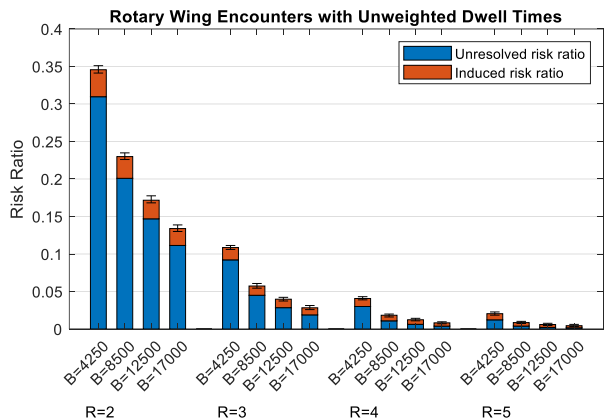

Figure 4. Rotary Wing Encounter Set with Unweighted Dwell Times

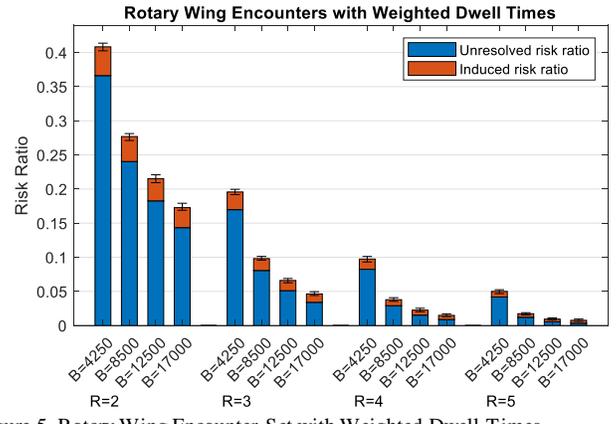

Figure 5. Rotary Wing Encounter Set with Weighted Dwell Times

*1) Dependence on Atmospheric Range*

The visual acquisition performance improved significantly with greater values for the atmospheric range, R. The search effectiveness had the biggest impact for lower R values. As R increased up to 5 miles, there was significantly less performance difference between different search effectiveness values. Thus, it was concluded that the model will be most sensitive to the atmospheric range, and that the pilot attentiveness scalar will be more important when the atmospheric range is small. We observed when comparing some configurations in the model, such as [$\beta = 4250$, R = 4] or [$\beta = 17000$, R = 3], that a good search effectiveness in poor atmospheric conditions can perform better than poor search effectiveness, in good visibility. Future flight test measurements and analysis will confirm if this behavior is accurate.

*2) DOV Dwell Time Impacts*

Figure 6 visualizes the comparison between fixed-wing single engine and rotorcraft encounters when using a weighted dwell distribution. Figure 6 does not decompose the results based on risk ratio component. By increasing the dwell time weights for the center left and center right DOV, there was priority to the pilot's central direction of view, which would be expected for a pilot's scanning technique. We hypothesized that while pilots scan forward more often than the sides, this in theory can result in fewer opportunities and less time to acquire due to faster closing rates across the FOV. For both encounter sets this resulted in a risk ratio increase.

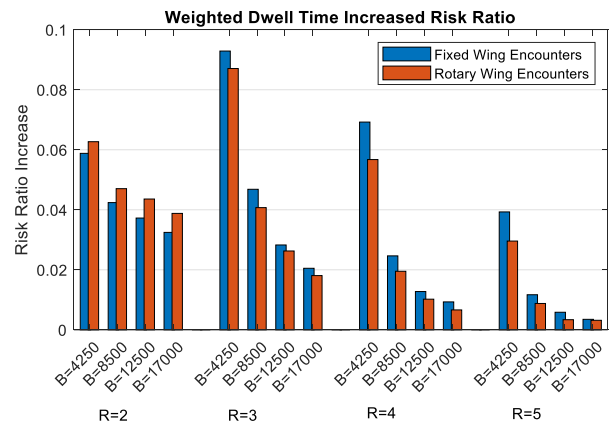

Figure 6. Increase in risk ratio when using weight dwell configuration compared to uniform dwell assumption.

*3) Fixed Wing vs Rotorcraft*

Figure 7 illustrates that the rotary wing encounter set had slightly less risk ratio then the fixed wing encounter set, but the differences between the two encounter sets decreased when visual acquisition parameters were increased. General trends were not dependent on if an unweighted or weighted dwell distribution. We hypothesize the increase is due to faster closing speeds for fixed wing encounters, resulting in less time to visually acquire and maneuver.

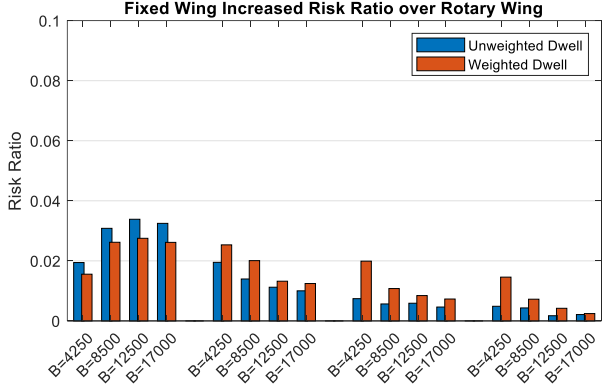

Figure 7. Increase in risk ratio of fixed wing over rotary wing encounters, organized by dwell configuration.

## VI. Discussion on Performance Requirements

The ASTM F3442/F3442M-20 DAA standard prescribes an NMAC risk ratio requirement of 0.30 when a drone encounters a crewed aircraft not equipped with ADS-B. This standard is applicable to the avoidance of crewed aircraft by a drone and implicitly assumes that only the drone will increase separation during a close encounter. Because of the small size of the drone, the crewed aircraft may not be able to visually acquire and avoid the drone. This standard assumes that a compliant DAA system would not be more hazardous during close encounters than current traditional operations. If the DAA NMAC performance requirements resulted in higher collision likelihoods than traditional operations, then the performance safety requirements would need to be changed.

Safety cases made to civil aviation authorities should estimate the likelihood of a MAC, which accounts for aircraft size and shapes; this can be calculated by multiplying the NMAC risk ratio requirement by a scaler for unmitigated P(MAC|NMAC). [19] estimated this to be 0.025 for see and be seen encounters (two traditional aircraft encounters) and in the range of [0.001, 0.0006] for drone vs traditional aircraft encounters. Like the ASTM standard, [19] assumed the largest applicable drone had a 25 feet wingspan but the largest prioritized wingspan was 13.7 feet. In [19], the wingspan of encountered aircraft ranged from 35.8 to 111 feet. If a DAA system satisfies the 0.30 NMAC risk ratio requirement, we can (NMAC Risk Ratio * P(MAC|NMAC)), to be within the set of [0.0003, 0.00018], shown by Equation 2, for the drone DAA performance standard.

$$0.30 \times \begin{bmatrix} 0.001 \\ 0.0006 \end{bmatrix} = \begin{bmatrix} 0.0003 \\ 0.00018 \end{bmatrix} \quad (2)$$

A wide range of "see and be seen" NMAC risk ratios, where both aircraft are responsible for maintain separation, were calculated under different conditions. For defining requirements, we propose selecting atmospheric range based on basic VFR weather minimums because, as discussed by Anderson et al.[24], "weather minimums serve as support for the pilot's duty of vigilance to see and avoid other aircraft;" and compliance to weather minimums is an enabler of see and avoid. Also, Perritt[25] argued that see and avoid is not possible when meteorological conditions obscure visibility. Specifically, 14 CFR § 91.155 prescribes as 3 miles or less for most operations. Figures 2-5 estimate the NMAC risk ratio to range [0.004, 0.24] and Equation 3 estimates see and be seen performance of two participating aircraft to be in the range of [0.0001, 0.006].

$$\begin{bmatrix} 0.004 \\ 0.24 \end{bmatrix} \times 0.025 = \begin{bmatrix} 0.0001 \\ 0.006 \end{bmatrix} \quad (3)$$

Given VFR weather minimums, the risk difference between the see and be seen performance of 0.0001 and the drone mitigations are on the order of $10^{-4}$ or less. We agree with Clothier et al. [11] and Santel [12] that the visual acquisition model overestimates human performance and we extend this conclusion in that using DAIDALUS for avoidance maneuvers also overestimates human performance. These results indicate that the ASTM F3442/F3442M-20 DAA risk ratio requirements will lead to aircraft interactions that are at least as safe as existing "see and be seen" interactions between crewed aircraft. The aviation community should review the use of weather minimums in estimating expected pilot performance. Better refinement of the "see and be seen" simulation, which could result in less conservative estimates of performance, could motivate ASTM to adopt less strict DAA risk ratio performance requirements.

## VII. Future Work

Various enhancements were made to a well-established model of visual acquisition and the model was integrated into an open-source simulation. However, these enhancements still result in a model that is overestimating human performance. Future work should prioritize refinement of the model to more accurately model human performance. This includes estimating a more representative search effectiveness value for general aviation based on upcoming flight test results funded by the FAA; improved modeling of scanning techniques; or assessing how detrimental aspects such as a dirty windshield, sunglasses, etc. that may reduce the clarity between the pilot and the target. could be used to tune the visual acquisition model. This future work could inform updates to the safety performance requirements in [2] and other international DAA standards.

## Acknowledgment

We like to thank our colleagues Matthew Edwards and Randal Guendel.